\documentclass{article}

\usepackage[T1]{fontenc}

\usepackage{amsmath,amsfonts}
\usepackage{amssymb}
\usepackage{graphicx}
\usepackage{verbatim}
\usepackage{mathrsfs}
\usepackage{url}
\usepackage{enumerate}
\usepackage{epstopdf}
\usepackage{amsthm}
\usepackage{thmtools}
\usepackage{thm-restate}
\usepackage{xcolor}

\newtheorem{definition}{Definition}
\newtheorem{lemma}{Lemma}
\newtheorem{corollary}{Corollary}
\newtheorem{theorem}{Theorem}
\newtheorem{claim}{Claim}
\newtheorem{question}{Question}

\newcommand{\pathsymbol}{\to}

\begin{document}

\title{Approximation Algorithms for the Bottleneck Asymmetric Traveling Salesman Problem
\footnotetext{A preliminary version of this work was presented in the 13th International Workshop on Approximation Algorithms for Combinatorial Optimization Problems and part of the thesis of the first author.}
}

\author{
Hyung-Chan An\thanks{Department of Computer Science, Yonsei University, Seoul 03722, South Korea. Research supported in part by NSF under grants no. CCR-0635121, DMS-0732196, CCF-0832782, CCF-0729102 and the Korea Foundation for Advanced Studies. Part of this research was conducted while the author was a PhD student at Cornell University. This work was supported by the National Research Foundation of Korea (NRF) grant funded by the Korea government (MSIT) (No. NRF-2019R1C1C1008934).} \and
Robert Kleinberg\thanks{Dept. of Computer Science, Cornell University, Ithaca, NY 14853. Supported by NSF grants no.\ CCF-0643934 and CCF-0729102, a grant from the Air Force Office of Scientific Research, a Microsoft Research New Faculty Fellowship, and an Alfred P.~Sloan Foundation Fellowship.} \and
David B. Shmoys\thanks{School of ORIE and Dept. of Computer Science, Cornell University, Ithaca, NY 14853. Research supported in part by NSF under grants no. CCR-0635121, DMS-0732196, CCF-0832782.}
}

\date{}

\maketitle

\begin{abstract}
We present the first nontrivial approximation algorithm for the {\it bottleneck asymmetric traveling salesman problem}. Given an asymmetric metric cost between $n$ vertices, the problem is to find a Hamiltonian cycle that minimizes its {\it bottleneck} (or maximum-length edge) cost. We achieve an $O(\log n / \log\log n)$ approximation performance guarantee by giving a novel algorithmic technique to shortcut Eulerian circuits while bounding the lengths of the shortcuts needed. This allows us to build on a related result of Asadpour, Goemans, M\k{a}dry, Oveis Gharan, and Saberi to obtain this guarantee. Furthermore, we show how our technique yields stronger approximation bounds in some cases, such as the bounded orientable genus case studied by Oveis Gharan and Saberi. We also explore the possibility of further improvement upon our main result through a comparison to the symmetric counterpart of the problem.
\end{abstract}

\section{Introduction}\label{s:int}

In this paper, we study the {\it bottleneck asymmetric traveling salesman problem}; that is, in contrast to the variant of the traveling salesman problem most commonly studied, the objective is to minimize the
maximum edge cost in the tour, rather than the sum of the edge costs. Furthermore, while the edge costs
satisfy the triangle inequality, we do not require that they be symmetric, in that the distance from
point $a$ to point $b$ might differ from the distance from $b$ to $a$. The triangle inequality is naturally satisfied by many cost functions; for example, minimizing the longest interval between job completions in the no-wait flow-shop reduces to the bottleneck asymmetric traveling salesman problem under a metric cost~\cite{P60,W72}. The bottleneck asymmetric traveling salesman problem cannot be approximated within a reasonable factor without assuming the triangle inequality. Surprisingly, no approximation algorithm was previously known to deliver solutions within an $o(n)$ factor of optimal, where $n$
denotes the number of nodes in the input. We present the first nontrivial approximation algorithm for the bottleneck asymmetric traveling salesman problem,
by giving an $O(\log n/ \log \log n)$-approximation algorithm. At the heart of our result is a new algorithmic technique for converting Eulerian circuits into tours while introducing ``shortcuts'' that are of bounded length.

For any optimization problem defined in terms of pairwise distances between nodes, it is natural 
to consider both the symmetric case and the asymmetric one, as well as the min-sum variant and
the bottleneck one. The standard (min-sum symmetric) traveling salesman problem (TSP) has been studied
extensively~\cite{LLRS}, and for approximation algorithms,
the 3/2-approximation algorithm of Christofides~\cite{C76} and, independently, Serdyukov~\cite{S78} remains the best known guarantee. 
The strongest NP-hardness result for this variant, due to
Karpinski, Lampis, and Schmied~\cite{KLS}, states that the existence of a $\rho$-approximation algorithm with $\rho < 123/122$ implies that P=NP. In contrast, for the bottleneck symmetric TSP, Lau~\cite{L81}, and Parker~\& Rardin~\cite{PR}, building on structural results of Fleischner~\cite{F74}, give a 2-approximation algorithm, and based on the metric in which all costs are either 1 or 2, it is easy to show that, for any $\rho <2$, the existence of a $\rho$-approximation algorithm implies that P=NP. For the asymmetric min-sum problem, Frieze, Galbiati, and Maffioli~\cite{FGM} gave the first $O(\log n)$-approximation algorithm, which is a guarantee that was subsequently matched by work of Kleinberg and Williamson~\cite{KW}, and 
improved upon by work of Asadpour, Goemans, M\k{a}dry, Oveis Gharan, and Saberi \cite{AGMOS}.
Most recently, in a breakthrough result, Svensson, Tarnawski, and V\'{e}gh~\cite{STV} gave the first constant approximation algorithm.

This cross-section of results is mirrored in other optimization settings. For example, for the 
min-sum symmetric $k$-median problem in which $k$ points are chosen as ``medians'' and each point is assigned to its nearest median, Arya, Garg, Khandekar, Meyerson, Munagala, and Pandit~\cite{AGKMMP} give a $\rho$-approximation algorithm for each $\rho >3$, whereas Jain, Mahdian, Markakis, Saberi, and Vazirani prove hardness results for $\rho < 1 + 2/e$~\cite{JMMSV}. In contrast, for the bottleneck symmetric version, the $k$-center problem, Hochbaum and Shmoys~\cite{HS} gave a 2-approximation algorithm, whereas Hsu and Nemhauser~\cite{HN} showed the NP-hardness of a performance guarantee of $\rho < 2$.
For the asymmetric $k$-center, a matching upper and lower bound of $\Theta(\log^* n)$ for the best performance guarantee was shown by Panigrahy \& Vishwanathan~\cite{PV98} and Chuzhoy, Guha, Halperin, Khanna, Kortsarz, Krauthgamer \& Naor~\cite{CGHKKKN}, respectively. In contrast, for the asymmetric
$k$-median problem, a bicriterion result 
which allowed a constant factor increase in cost with a logarithmic increase 
in the number of medians was shown by Lin and Vitter~\cite{LV}, and a hardness tradeoff matching this (up to constant factors) was proved by Archer~\cite{A00}.

In considering these comparative results, there is a mixed message as to whether a bottleneck problem is easier or harder to approximate than its min-sum counterpart. On the one hand, for any bottleneck problem, one can immediately reduce the optimization problem with cost data to a more combinatorially
defined question, since there is the trivial relationship that the optimal bottleneck solution is of objective function value at most $T$ if and only if there exists a feasible solution that uses only those edges of cost at most $T$. Furthermore, there are only a polynomial number of potential
thresholds $T$, and so a polynomial-time algorithm that answers this purely combinatorial decision question leads to a polynomial-time optimization algorithm.

Similarly, for a $\rho$-approximation algorithm, it
is sufficient for the algorithm to solve a ``relaxed'' decision question: either provide some certificate that no feasible solution exists, or produce a solution in which each edge used is of cost at most $\rho T$.
If $G$ denotes the graph of all edges of cost at most $T$, then the triangle inequality implies that
it is sufficient to find feasible solutions within $G^{\rho}$, the $\rho$th power of $G$, in which we include an
edge $(u,v)$ whenever $G$ contains a path from $u$ to $v$ with at most $\rho$ edges. In the context
of the TSP, this means that we either want to prove that $G$ is not Hamiltonian, or else to produce
a Hamiltonian cycle within, for example, the square of $G$ (to yield a 2-approximation algorithm as in~\cite{L81,PR}).

Unfortunately, the techniques invented in the context of the min-sum problem do not seem to be amenable to bottleneck objective function. For example, the analysis of the $O(\log n)$-approximation algorithm for the min-sum asymmetric TSP due to Kleinberg and Williamson~\cite{KW} depends crucially on the monotonicity of the optimal value over the vertex-induced subgraphs, and the fact that shortcutting a circuit does not increase the objective. That fact clearly is not true in the bottleneck setting: shortcutting arbitrary subpaths of a circuit may result in a tour that is valid only in a higher-order power graph. The aforementioned monotonicity is also lost as it relies on this fact as well.

In order to resolve this difficulty, we devise a condition on Eulerian circuits under which we can limit the lengths of the paths that are shortcut to obtain a Hamiltonian cycle. We will present a polynomial-time constructive proof of this condition using Hall's Transversal Theorem~\cite{H35}; this proof is directly used in the algorithm. One of the special cases of the condition particularly worth mentioning is a degree-bounded spanning circuit (equivalently, an Eulerian spanning subgraph of bounded degree). If there exists a bound $k$ on the number of occurrences of any vertex in a spanning circuit, our theorem provides a bound of $2k-1$ on the length of the shortcut paths.

We will then show how thin trees defined in Asadpour et al.~\cite{AGMOS} can be used to compute these degree-bounded spanning circuits. A $\beta$-thin tree with respect to a weighted graph $G$ is a unit-weighted spanning tree of $G$ whose cut weights are no more than $\beta$ times the corresponding cut weights of $G$. The min-sum algorithm due to 
Asadpour et al.~\cite{AGMOS} augments an $O(\log n/\log\log n)$-thin tree with respect to a (scaled) Held-Karp solution (Held and Karp~\cite{HK}) into a spanning Eulerian graph by solving a circulation problem. The Held-Karp relaxation is a linear program consisting of the equality constraints on the in- and out-degree of each vertex and the inequality constraints on the directed cut weights: the equality constraints set the degrees to one, and the inequality constraints ensure that the total weight of edges leaving $S$ is at least 1 for each subset $S$. We introduce vertex capacities to the circulation problem to impose the desired degree bound without breaking the feasibility of the circulation problem. This leads to an algorithm that computes degree-bounded spanning circuits with an $O(\log n/\log\log n)$ bound.

Oveis Gharan and Saberi~\cite{OS} subsequently gave an $O(1)$-approximation algorithm for the min-sum asymmetric TSP when the support of the Held-Karp solution can be embedded on an orientable surface with a bounded genus. They achieved this by showing how to extract an $O(1)$-thin tree in this special case. Our result can be combined with this result to yield an $O(1)$-approximation algorithm for the bottleneck asymmetric TSP when the support of the Held-Karp solution has a bounded orientable genus. Chekuri, Vondr{\'a}k, and Zenklusen~\cite{CVZ} showed that an alternative sampling procedure can be used to find the thin tree in Asadpour et al.~\cite{AGMOS}.

Section~\ref{s:pre} of this paper reviews some notation and the needed details of previous results, and Section~\ref{s:alg} describes the $O(\log n/\log\log n)$-approximation algorithm to the bottleneck asymmetric traveling salesman problem. Section~\ref{s:sc} examines the special case when the support of the Held-Karp solution can be embedded on an orientable surface with a bounded genus. Some open questions are discussed in Section~\ref{s:opd}.

As a final remark and, as mentioned above, subsequent to the appearance of a preliminary version of this paper, Svensson et al.~\cite{STV} gave the first $O(1)$-approximation algorithm for the min-sum asymmetric TSP; their approximation ratio was later improved by Traub and Vygen~\cite{TV}. 
The question of whether the new methods in these papers can be adapted to yield similar improvements for the bottleneck asymmetric TSP remains an enticing open problem.

\section{Preliminaries}\label{s:pre}

We introduce some notation and review previous results in this section. Some notation was adopted from Asadpour et al.~\cite{AGMOS}.

Let $G=(V,A)$ be a digraph and $E$ be the underlying undirected edge set: $\{u,v\}\in E$ if and only if $\langle u,v\rangle \in A$ or $\langle v,u\rangle \in A$. For $S\subsetneq V$, let\begin{eqnarray*}
\delta^+(S) &:=& \{\langle u,v\rangle\in A\mid u\in S,v\notin S\},\\
\delta^-(S) &:=& \delta^+ (V\setminus S),\\
\delta(S) &:=& \{ \{u,v\}\in E\mid |E\cap S|=1\};
\end{eqnarray*}for $v\in V$,\begin{eqnarray*}
\delta^+(v) &:=& \delta^+(\{v\}),\\
\delta^-(v) &:=& \delta^-(\{v\}),\\
\delta(v) &:=& \delta(\{v\}).
\end{eqnarray*}

Given a vector $z\in\mathbb{R}^E$, the \emph{support} of $z$ is defined to be the set of edges on which $z$ has a nonzero value, i.e., $\{e\mid z_e\neq 0\}$.
For $z\in\mathbb{R}^E$ and $F\subseteq E$, let $z(F)$ be a shorthand for $z(F) := \sum_{f\in F} z_f$. Similarly, for $x\in\mathbb{R}^A$ and $B\subseteq A$, let $
x(B) := \sum_{b\in B} x_b$.

We need a notion of the non-Hamiltonicity certificate to solve the ``relaxed'' decision problem. We establish this certificate by solving the Held-Karp relaxation~\cite{HK} in our algorithm. The Held-Karp relaxation to the asymmetric traveling salesman problem is the following linear program. We do not define an objective here.\begin{equation}
\begin{cases}\label{e:hk}
x(\delta^+(v))=x(\delta^-(v))=1, &\forall v\in V;\\
x(\delta^+(S))\geq 1 ,&\forall S\subsetneq V,S\neq \emptyset;\\
x\geq 0 .&
\end{cases}
\end{equation}A graph is non-Hamiltonian if \eqref{e:hk} is infeasible. This linear program can be solved in polynomial time~\cite{GLS}.

A \emph{thin tree} is defined as follows in Asadpour et al.~\cite{AGMOS}.
\begin{definition}
A spanning tree $T$ is \emph{$\beta$-thin} with respect to $z^*\in \mathbb{R}_+^E$ if \linebreak
$|T\cap \delta(U)|\leq\beta z^*(\delta(U))$ for all $U\subsetneq V$.
\end{definition}
\noindent In particular, we will call a thin tree that is a subset of the support of $z^*$ a \emph{supported thin tree}.
\begin{definition}
A tree $T$ is called a \emph{supported $\beta$-thin tree} with respect to $z^*$ if $T$ is $\beta$-thin with respect to $z^*$ and every edge of $T$ is in the support of $z^*$.
\end{definition}
\noindent Asadpour et al.~\cite{AGMOS} prove Theorem~\ref{t:thin}: they show the thinness for $z^*_{uv} := \frac{n-1}{n} ( x^*_{uv} + x^*_{vu} )$ where $n=|V|$, and Theorem~\ref{t:thin} is slightly weaker.
\begin{theorem}
\label{t:thin}
There exists a probabilistic algorithm that, given an extreme point solution $x^*\in\mathbb{R}^A$ to the Held-Karp relaxation, produces a supported $\beta$-thin tree $T$ with respect to $z^*_{uv} := x^*_{uv} + x^*_{vu}$ with high probability, for $\beta=\frac{4\ln n}{\ln\ln n}$.
\end{theorem}

Let $T_\to$ be a directed version of $T$, obtained by choosing the arcs in the support of $x^*$. If arcs exist in both directions, an arbitrary choice can be made. We shall consider a circulation problem on $G$: recall that the circulation problem requires, given a lower and upper bound on each arc, a set of flow values on arcs such that the sum of the incoming flows at every vertex matches the sum of outgoing, while honoring both bounds imposed on each arc. When all of the bounds are integers, an integral solution can be found in polynomial time unless the problem is infeasible~\cite{S03}. Here we consider an instance where the lower bounds $\ell$ and upper bounds $u$ on the arcs are given as follows:\begin{equation}
\label{e:circ}
\begin{split}
\ell(a)&=\begin{cases}1,&\textrm{if }a\in T_\to ,\\0,&\textrm{otherwise},\end{cases}\\
u(a)&=\begin{cases}1+2\beta x_a^*,&\textrm{if }a\in T_\to ,\\2\beta x_a^* ,&\textrm{otherwise}.\end{cases}
\end{split}
\end{equation}
Asadpour et al.~\cite{AGMOS} show that this problem is feasible; the existence of an integral circulation under the rounded-up bounds $\lceil u(a)\rceil$ for each $a\in A$ follows from that.
\begin{lemma}
\label{l:orgfeas}
The circulation problem defined by \eqref{e:circ} is feasible.
\end{lemma}

\section{Algorithm}\label{s:alg}

This section gives the $O(\frac{\log n}{\log\log n})$-approximation algorithm for the bottleneck asymmetric traveling salesman problem and its analysis. We present the lemmas to bound the lengths of the paths that are shortcut in the process of transforming a spanning circuit into a Hamiltonian cycle; we also show how a degree-bounded spanning circuit can be constructed. Throughout this paper, we shall use the term \emph{circuit} to emphasize the fact that a spanning circuit is allowed to visit the same vertex more than once, as opposed to a Hamiltoninan cycle that has to visit every vertex exactly once.

\begin{lemma}
\label{l:cond}
Let $v_1,\ldots,v_m,v_1$ be a (non-simple) circuit that visits every vertex at least once. Partition $v_1,\ldots,v_m$ into contiguous subsequences of length $k$, except for the final subsequence whose length may be less than $k$. Denote the pieces of this partition by $P_1,\ldots,P_\ell$. If, for all $t$, the union of any $t$ sets in $\{P_1,\ldots,P_\ell\}$ contains at least $t$ distinct vertices, $G^{2k-1}$ is Hamiltonian.
\end{lemma}
\begin{proof}
From Hall's Transversal Theorem~\cite{H35}, if the given condition holds,\linebreak $\{P_1,\ldots,P_\ell\}$ has a transversal: i.e., we can choose one vertex from each piece $P_i$ such that no vertex is chosen more than once. If we take any subsequence of $v_1,\ldots,v_m$ that contains every vertex exactly once and includes all of the vertices in the transversal, this subsequence is a Hamiltonian cycle in $G^{2k-1}$. This is because any two contiguous vertices chosen in the transversal are at most $2k-1$ arcs away. Since a transversal can be found in polynomial time (see Kleinberg and Tardos~\cite{KT}), a Hamiltonian cycle can be constructed in polynomial time as well.
\end{proof}

Lemma~\ref{l:bdd} shows that a degree-bounded spanning circuit forms a special case of Lemma~\ref{l:cond}.
\begin{lemma}
\label{l:bdd}
Given a circuit on $G$ that visits every vertex at least once and at most $k$ times, a Hamiltonian cycle on $G^{2k-1}$ can be found in polynomial time.
\end{lemma}
\begin{proof}
Consider $\{P_1,\ldots,P_\ell\}$ as defined in Lemma~\ref{l:cond}. For any $t$ sets in $\{P_1,\ldots,P_\ell\}$, the sum of their cardinalities is strictly greater than $(t-1)k$. If their union contained only $t-1$ distinct vertices, then by the pigeonhole principle there would be some vertex that occurs at least $k+1$ times, violating the upper bound on the number of occurrences of any vertex in the circuit.

Thus, by Lemma~\ref{l:cond}, there exists a Hamiltonian cycle in $G^{2k-1}$, and this can be found in polynomial time.
\end{proof}

Now we show how to construct a degree-bounded spanning circuit.
\begin{lemma}
\label{l:newfeas}
Let $x^*$ be a feasible solution to the Held-Karp relaxation. Given a supported $\beta$-thin tree $T$ with respect to $z^*_{uv} := x^*_{uv} + x^*_{vu}$, a circuit on $G$ with every vertex visited at least once and at most $\lceil 4\beta \rceil$ times can be found in polynomial time.
\end{lemma}
\begin{proof}
We modify the circulation problem defined in \eqref{e:circ} by introducing vertex capacities to the vertices: every vertex $v$ is split into two vertices $v_i$ and $v_o$, where all the incoming edges are connected to $v_i$ and the outgoing edges are from $v_o$. We set the vertex capacity $u(\langle v_i,v_o\rangle)$ as $\sum_{a:\mathrm{tail}(a)=v}u(a)$. (See Fig.~\ref{f:1}.) It is easy to see that this modification does not change the feasibility; thus, from Lemma~\ref{l:orgfeas}, this new circulation problem instance is also feasible.
\begin{figure}
\centering
\includegraphics[width=\textwidth]{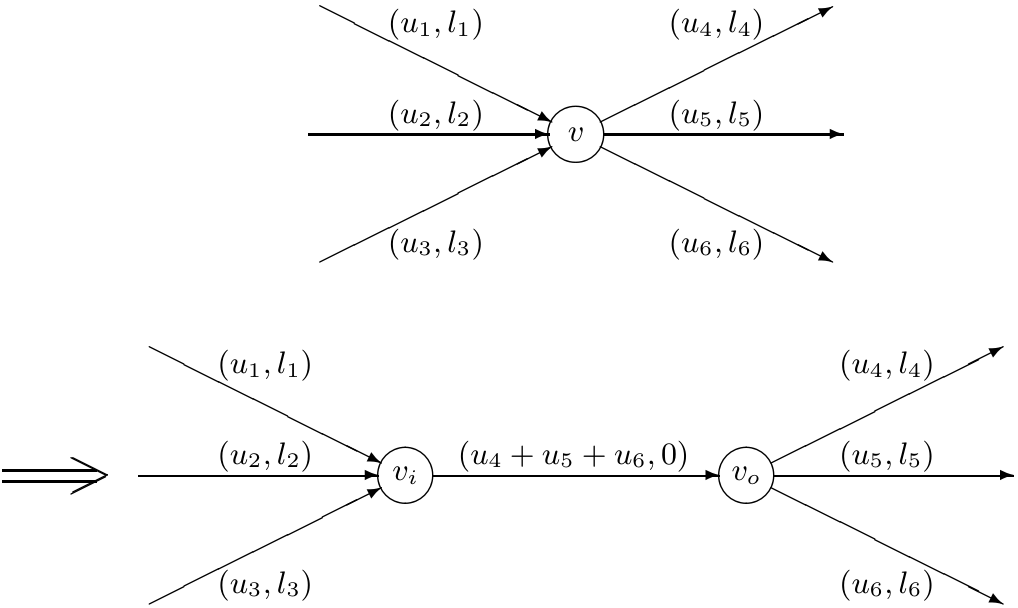}
\caption{Introducing vertex capacities.}
\label{f:1}
\end{figure}

Rounding up all $u$ values of this instance preserves the feasibility and guarantees the existence of an integral solution. By contracting split vertices back in the integral solution, we obtain a spanning Eulerian subgraph of $G=(V,A)$ (with arcs duplicated) whose maximum indegree is at most $\max_{v\in V}\lceil \sum_{a:\mathrm{tail}(a)=v}u(a) \rceil$. Observe that, for any $v\in V$,\begin{eqnarray*}
\sum_{a:\mathrm{tail}(a)=v}u(a) &=& |\{a\in T_\to \mid \mathrm{tail}(a)=v\}| + \sum_{a:\mathrm{tail}(a)=v}2\beta x_a^*\\
&\leq& \beta z^*(\delta(v)) + 2\beta x^*(\delta^+(v))\\
&=&4\beta
.\end{eqnarray*}Thus, we can find a spanning Eulerian subgraph of $G=(V,A)$ whose maximum degree is at most $\lceil 4\beta\rceil$, given the $\beta$-thin tree $T$. Any Eulerian circuit of this graph will satisfy the desired property.
\end{proof}

Note that it is important to introduce these vertex capacities since the maximum degree of the support of an extreme point Held-Karp solution can grow linearly in the number of vertices. In fact, both the indegree and outdegree of the same vertex can grow linearly. This would give $\sum_{a:\mathrm{tail}(a)=v}\lceil u(a)\rceil = \Theta(n)$ for such vertex $v$. (See Theorem~\ref{t:lindegextpt} in Appendix~\ref{app:deg}.)

Theorem~\ref{t:thin} and Lemmas~\ref{l:bdd} and \ref{l:newfeas} yield the algorithm.
\begin{theorem}
\label{t:main}
There exists a probabilistic $O(\frac{\log n}{\log \log n})$-approximation algorithm for the bottleneck asymmetric traveling salesman problem under a metric cost.
\end{theorem}
\begin{proof}
Let $A_{\leq\tau}:=\{\langle u,v\rangle\mid c(u,v)\leq\tau\}$ and $G_{\leq\tau}:=(V,A_{\leq\tau})$. The algorithm first determines the minimum $\tau$ such that the Held-Karp relaxation for $G_{\leq\tau}$ is feasible. Let $\tau^*$ be this minimum. If $\tau_1\leq\tau_2$ and the Held-Karp relaxation for $G_{\leq\tau_2}$ is infeasible, the relaxation for $G_{\leq\tau_1}$ is also infeasible; therefore, $\tau^*$ can be discovered by binary search. Note that $\tau^*$ can serve as a lower bound on the optimal solution value.

Once $\tau^*$ is determined, we compute an extreme point solution $x^*$ to the Held-Karp relaxation for $G_{\leq\tau^*}$. Then we sample a $\beta$-thin tree $T$ with respect to $z^*_{uv} := x^*_{uv} + x^*_{vu}$ for $\beta=\frac{4\ln n}{\ln\ln n}$. By Theorem~\ref{t:thin}, this can be performed in polynomial time with high probability.

Then the algorithm constructs the circulation problem instance described in the proof of Lemma~\ref{l:newfeas} and finds an integral solution. Lemma~\ref{l:newfeas} shows that any Eulerian circuit of this integral solution is a spanning circuit where no vertex appears more than $\lceil 4\beta\rceil$ times.

Let $\{P_1,\ldots,P_\ell\}$ be the partition of this spanning circuit as defined in Lemma~\ref{l:cond} for $k=\lceil 4\beta\rceil$. The algorithm computes a transversal of $\{P_1,\ldots,P_\ell\}$ and augments it into a Hamiltonian cycle $C$ in $G^{2\lceil 4\beta\rceil -1}$. By the triangle inequality, the cost of $C$ is at most $(2\lceil 4\beta\rceil -1)\cdot\tau^*$; thus, $C$ is a $(2\lceil 4\beta\rceil -1)$-approximate solution to the given input. Note that $2\lceil 4\beta\rceil -1 = 2\lceil \frac{16\ln n}{\ln\ln n}\rceil - 1 = O(\frac{\log n}{\log\log n})$.

The foregoing is a probabilistic $O(\frac{\log n}{\log \log n})$-approximation algorithm for the bottleneck asymmetric traveling salesman problem under a metric cost.
\end{proof}

\section{Improved Bound for Constant Genus Support Graphs}\label{s:sc}

In this section, we illustrate how our framework can be used together with other results to yield a stronger approximation guarantee in certain special cases. Lemmas~\ref{l:bdd} and \ref{l:newfeas} imply the following theorem.
\begin{theorem}
\label{t:sc}
If a supported $f(n)$-thin tree can be found in polynomial time for a certain class of metric, an $O(f(n))$-approximation algorithm exists for the bottleneck asymmetric traveling salesman problem under the same class of metric.
\end{theorem}

In particular, Oveis Gharan and Saberi~\cite{OS} investigate the case when the Held-Karp solution can be embedded on an orientable surface with a bounded genus; Oveis Gharan and Saberi~\cite{OS}, in addition to an $O(1)$-approximation algorithm for the min-sum problem, show the following:
\begin{theorem}
\label{t:bgthin}
Given a feasible solution $x^*\in\mathbb{R}^A$ to the Held-Karp relaxation, let $z^*_{uv} := x^*_{uv} + x^*_{vu}$. If the support of $z^*$ can be embedded on an orientable surface with a bounded genus, a supported $\beta$-thin tree with respect to $z^*$ can be found in polynomial time, where $\beta$ is a constant that depends on the bound on the genus.
\end{theorem}
Theorems~\ref{t:sc} and \ref{t:bgthin} together imply the following.
\begin{corollary}
\label{c:bg}
There exists an $O(1)$-approximation algorithm for the bottleneck asymmetric traveling salesman problem when the support of the Held-Karp solution can be embedded on an orientable surface with a bounded genus.
\end{corollary}

\section{Open Problems and Discussion}\label{s:opd}

Given that the bottleneck symmetric TSP is 2-approximable~\cite{F74,L81,PR}, a naturally following question is if the asymmetric version also admits a 2-approximation algorithm. The algorithms for the symmetric case are based on the fact that the square of a 2-connected graph is Hamiltonian. One could regard the analogue of 2-connectedness of an undirected graph in a digraph as the following property: for any two vertices, there exists a simple directed cycle that includes both vertices. However, unfortunately, there exists such a graph whose square is non-Hamiltonian. In fact, for any constant $k$ and $p$, the following holds.
(The proof of this theorem is deferred to the end of this section.)
\begin{restatable}{theorem}{restatevdnh}
\label{t:vdnh}
For any constant $k,p\in\mathbb{N}$, there exists a digraph $G=(V,A)$ such that:\begin{quote}\begin{enumerate}[(i)]\itemsep0.5ex
\item for all $u,v\in V$, there exist $k$ paths $P_1,\ldots,P_k$ from $u$ to $v$ and $k$ paths $Q_1,\ldots,Q_k$ from $v$ to $u$ such that $P_1,\ldots,P_k,Q_1,\ldots,Q_k$ are internally vertex-disjoint;\label{i:t:vdnh:1}
\item $G^p$ is non-Hamiltonian.\label{i:t:vdnh:2}
\end{enumerate}\end{quote}
\end{restatable}
\noindent As this approach appears unpromising, one could instead ask if some constant-order power of a graph whose Held-Karp relaxation is feasible is Hamiltonian.

\begin{question}
\label{q:oq}
Does there exist a constant $p$ such that the $p$th power of any digraph with a feasible Held-Karp relaxation is Hamiltonian?
\end{question}

One plausible way to affirmatively answer Question~\ref{q:oq} is by proving that a graph whose Held-Karp relaxation is feasible contains a spanning circuit that satisfies the property of Lemma~\ref{l:cond}; Lemma~\ref{l:bdd} might be helpful in this. In particular, if there exists an efficient procedure that computes an $O(1)$-thin tree with respect to the Held-Karp solution, that would yield an $O(1)$-approximation algorithm for the bottleneck asymmetric TSP.

Considering the undirected case, we can show that the set of graphs whose Held-Karp relaxation is feasible is a proper subset of the set of 2-connected graphs (see Theorem~\ref{t:shk2} for one direction); given this observation, it is conceivable that one could attain a direct and simpler proof that the square of a graph whose Held-Karp relaxation is feasible is Hamiltonian. Such a proof may provide inspiration for the asymmetric case.

\begin{theorem}
\label{t:shk2}
For an undirected graph $G=(V,E)$, if the linear system\begin{equation}\label{e:shk}
\begin{cases}
z(\delta(v))=2 ,&\forall v\in V;\\
z(\delta(S))\geq 2 ,&\forall S\subsetneq V,S\neq \emptyset;\\
z\geq 0 &
\end{cases}
\end{equation}has a feasible solution $z^*\in\mathbb{R}^E$, $G$ is 2-connected.
\end{theorem}

We conclude this section with the proofs of the two theorems. Theorem~\ref{t:shk2} was previously shown by Boyd and Elliott-Magwood~\cite{BE07,BE10}; however, for the sake of completeness, we present a short proof that is based on the ideas from the proof of Lemma~\ref{l:newfeas}.

\begin{proof}[Proof of Theorem~\ref{t:shk2}]
Let $G'=(V,A)$ be the digraph obtained from $G$ by replacing each edge with two arcs in both directions. Consider a flow network on $G'$, where the arc capacity is given as the $z^*$ value of the underlying edge.

For any $u,v\in V$, a flow of 2 can be routed from $u$ to $v$ on this network. Let $f\in\mathbb{R}^A$ be this flow. Without loss of generality, we can assume that\begin{equation}\label{e:t:shk2:1}
\forall\{x,y\}\in E\quad f(x,y)=0\textrm{ or }f(y,x)=0
.\end{equation}
We drop the arcs on which the flow is zero from the network.

Let $x$ be an arbitrary vertex other than $u$ or $v$. Note that, from \eqref{e:t:shk2:1}, the sum of the capacities of the arcs incident to/from $x$ is at most 2. From the flow conservation, the incoming flow into $x$ is at most 1; thus, introducing the vertex capacity of 1 to every vertex other than $u$ and $v$ does not break the feasibility of $f$.

Now we round up all of the capacities, and there exists an integral flow of value 2 from $u$ to $v$ on this flow network. This proves the existence of two vertex-disjoint paths from $u$ to $v$.
\end{proof}

\begin{proof}[Proof of Theorem~\ref{t:vdnh}]
We construct a digraph $G=(V,A)$ that satisfies the desired properties as follows. (See Figure~\ref{f:2}.)
\begin{itemize}
\item $V$ is given as the union of $p+1$ disjoint sets: $V:=\cup_{i\in\mathbb{F}_{p+1}} V_i$. These sets satisfy $|V_0|=p(2k+2)+1$ and $|V_1|=\cdots=|V_p|=2k+2$. We denote the vertices in $V_i$ as $v_{i,1},\ldots,v_{i,|V_i|}$.
\item For all $i$, there is an arc from every vertex in $V_i$ to every vertex in $V_{i+1}$: i.e., the subgraph induced by $V_i\cup V_{i+1}$ is a complete bipartite digraph. No other arcs exist in the graph.
\end{itemize}

\begin{figure}[t]
\centering
\includegraphics[width=.9\textwidth]{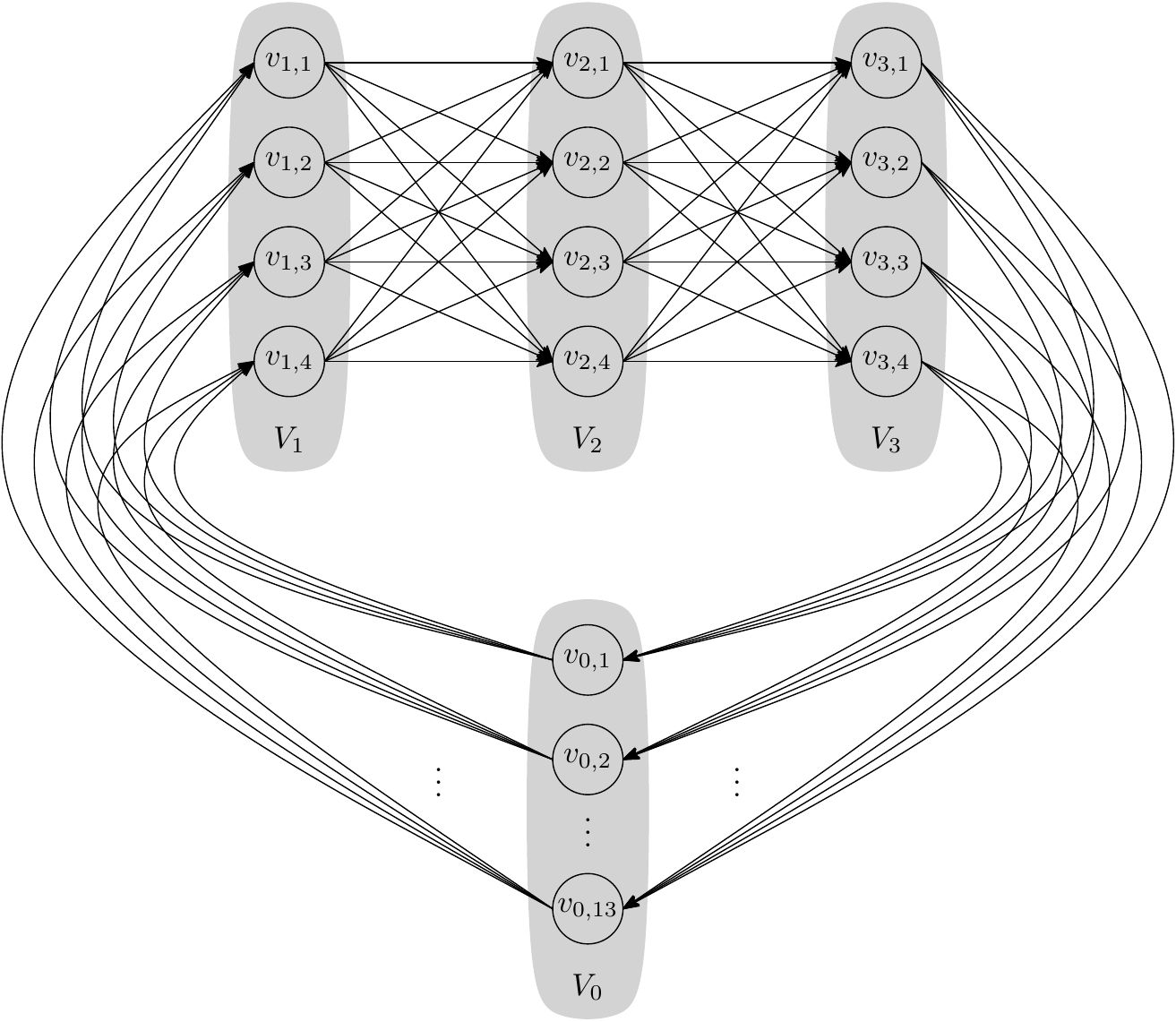}
\caption{Construction for $k=1$, $p=3$.}
\label{f:2}
\end{figure}

First we verify Property~(\ref{i:t:vdnh:1}). Let $v_{i_1,j_1}$ and $v_{i_2,j_2}$ be two arbitrary vertices in $G$. From the construction, we can choose two sets $\mathcal{P}$ and $\mathcal{Q}$ such that $\mathcal{P}, \mathcal{Q}\subseteq\{1,\ldots,2k+2\}\setminus\{j_1,j_2\}$, $|\mathcal{P}|=|\mathcal{Q}|=k$, and $\mathcal{P}\cap\mathcal{Q}=\emptyset$. Now for each $j^*\in\mathcal{P}$, consider the path $v_{i_1,j_1} \pathsymbol v_{i_1+1,j^*} \pathsymbol \cdots \pathsymbol v_{i_2-1,j^*} \pathsymbol v_{i_2,j_2}$. This path is three vertices long when $i_1+2=i_2$ and $p+3$ vertices long when $i_1+1=i_2$. Similarly, for each $j^*\in\mathcal{Q}$, consider the path $v_{i_2,j_2} \pathsymbol v_{i_2+1,j^*} \pathsymbol \cdots \pathsymbol v_{i_1-1,j^*} \pathsymbol v_{i_1,j_1}$. Note that these $2k$ paths are internally vertex-disjoint.

Now we verify Property~(\ref{i:t:vdnh:2}). Suppose $G^p$ has a Hamiltonian cycle. Then we can associate each $v\in V_0$ with the vertex $u$ that follows $v$ in the Hamiltonian cycle. Observe that $u$ is adjacent from $v$ and that these associated vertices are distinct. On the other hand, in $G^p$, the set of vertices that are adjacent from a vertex in $V_0$ is exactly equal to $V_1\cup\cdots\cup V_p$ from the construction. Note that $|V_0|>|V_1\cup\cdots\cup V_p|$; hence, by the pigeonhole principle, $G^p$ is not Hamiltonian.
\end{proof}

\appendix
\section*{Appendix}

\section{Maximum degree of extreme-point Held-Karp solutions}\label{app:deg}

In this appendix, we show that the maximum degree of the support of an extreme point Held-Karp solution can grow linearly in the number of vertices, by providing an explicit construction of such extreme point solutions.

\begin{theorem}\label{t:lindegextpt}
For any $k\geq 2$, the Held-Karp relaxation on $2k+5$ vertices has an extreme point solution whose support contains a vertex with indegree $k$ and outdegree $k$.
\end{theorem}
\begin{proof}
Let $V:=\{u_0,\ldots,u_{k+1},v_0,\ldots,v_{k+1},w\}$ be the set of $2k+5$ vertices. We define an extreme point Held-Karp solution $x^*\in\mathbb{R}_+^{V\times V}$ as follows: Figure~\ref{f:3} illustrates the $k=4$ case.
\begin{enumerate}[(i)]
\item $x^*(v_{k+1},u_0)=1$;\label{i:t:lindegextpt:x1}
\item for all $0\leq i\leq k$, $x^*(u_i,u_{i+1})=x^*(v_i,v_{i+1})=1-\frac{1}{k}$;\label{i:t:lindegextpt:x2}
\item for all $1\leq i\leq k$, $x^*(u_{k+1},u_i)=x^*(u_i,w)=x^*(w,v_i)=x^*(v_i,v_0)=\frac{1}{k}$;\label{i:t:lindegextpt:x3}
\item $x^*(u_0,u_{k+1})=x^*(v_0,v_{k+1})=\frac{1}{k}$.\label{i:t:lindegextpt:x4}
\end{enumerate}

\begin{figure}[t]
\centering
\includegraphics[width=\textwidth]{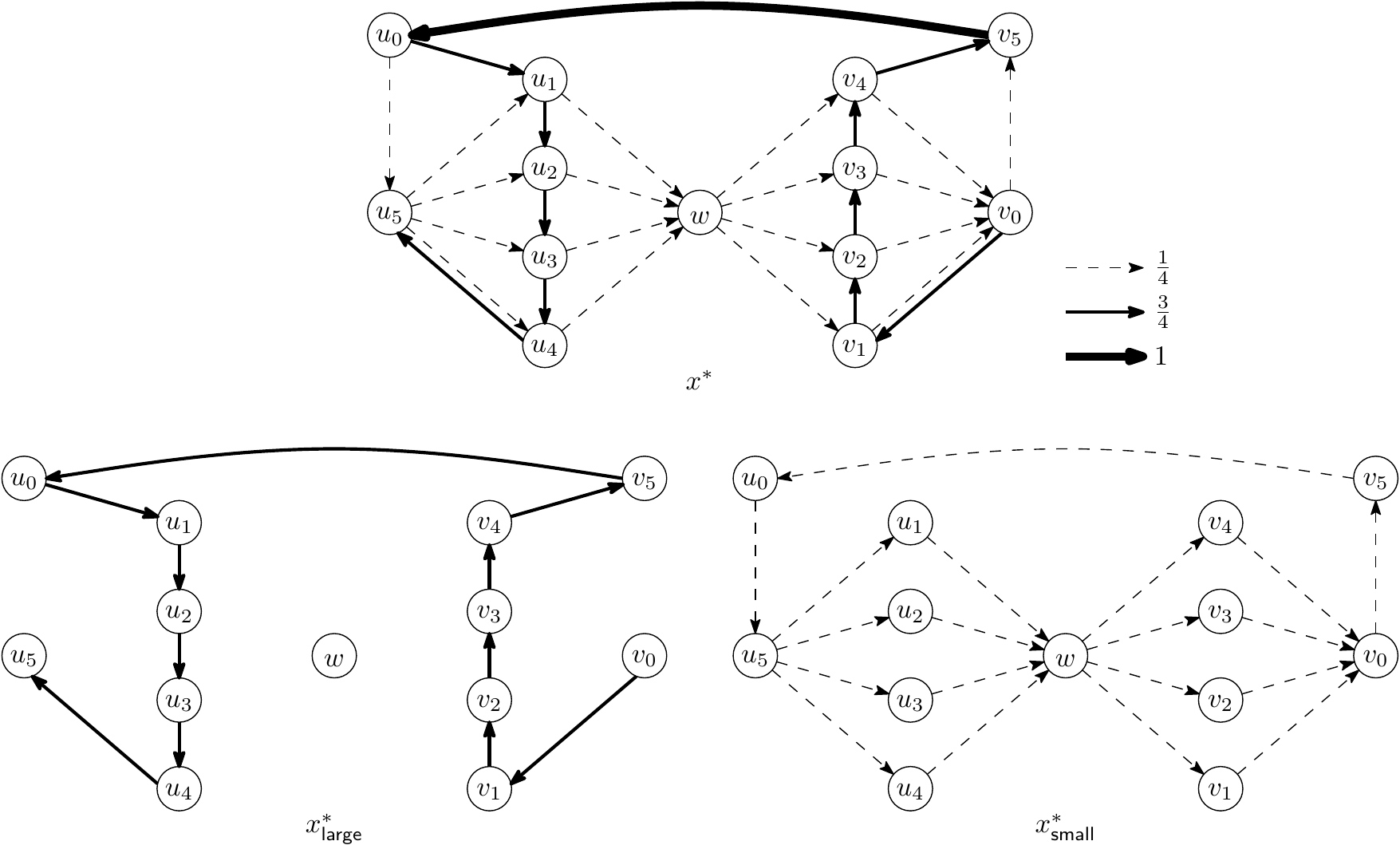}
\caption{An extreme point solution $x^*$ when $k=4$, and its decomposition into $x^*_{\mathsf{large}}$ and $x^*_{\mathsf{small}}$.}
\label{f:3}
\end{figure}
\noindent We note that the left (or right) half of this construction has a similar structure to a Held-Karp solution that was given by Elliott-Magwood~\cite[Theorem~6.4.1]{E08} to determine the integrality gap of the relaxation. That solution, however, cannot be directly used here because we need a vertex whose indegree and outdegree are both large. (The solution also is an extreme point solution although this fact was not observed there.)

We will first prove that $x^*$ is a feasible solution to the Held-Karp relaxation; then, we will choose a set of LP constraints that are tightly satisfied by $x^*$ and show that the linear system of these constraints (where inequalities are replaced by equalities) does not have two distinct solutions. This proves that $x^*$ is an extreme point solution.

\begin{claim}
$x^*$ is a feasible Held-Karp solution.
\end{claim}
\begin{proof}
It can be easily verified that $x^*(\delta^+(v))=x^*(\delta^-(v))=1$ for all $v$; it remains to verify that $x^*(\delta^+(S))\geq 1$ for all nonempty $S\subsetneq V$.

We decompose $x^*$ into the sum of two vectors $x^*_{\mathsf{large}},x^*_{\mathsf{small}}\in\mathbb{R}_+^{V\times V}$, where every nonzero arc of $x^*_{\mathsf{large}}$ has the value of $1-\frac{1}{k}$ and every nonzero arc of $x^*_{\mathsf{small}}$ has $\frac{1}{k}$. $x^*_{\mathsf{large}}$ is supported by the arcs whose $x^*$ value is defined in Cases~(\ref{i:t:lindegextpt:x1}) and (\ref{i:t:lindegextpt:x2}); $x^*_{\mathsf{small}}$ is supported by the arcs defined in Cases~(\ref{i:t:lindegextpt:x1}), (\ref{i:t:lindegextpt:x3}), and (\ref{i:t:lindegextpt:x4}): $\langle v_{k+1},u_0\rangle$ is the only arc that belongs to both supports. It is trivial to see $x^*=x^*_{\mathsf{large}}+x^*_{\mathsf{small}}$.

Since the support of $x^*_{\mathsf{small}}$ is strongly connected, we have $x^*_{\mathsf{small}}(\delta^+(S))\geq\frac{1}{k}$; hence, if the support of $x^*_{\mathsf{large}}$ has at least one edge in $\delta^+(S)$, $x^*(\delta^+(S))\geq 1$. Suppose from now that the support of $x^*_{\mathsf{large}}$ does not have any edge in $\delta^+(S)$. As the support of $x^*_{\mathsf{large}}$ forms a path that visits every vertex other than $w$, one of the following holds: (\textbf{a}) $S=\{w\}$, (\textbf{b}) $S=V\setminus \{w\}$, or (\textbf{c}) $S$ is given as a nonempty strict postfix of the path $v_0\to\cdots\to v_{k+1}\to u_0\to\cdots\to u_{k+1}$. It is easy to verify that $x^*(\delta^+(S))=x^*_{\mathsf{small}}(\delta^+(S))=1$ in the first two cases; in the last case, we have $u_{k+1}\in S$ and $v_0\notin S$. If $w\notin S$, either $\langle u_{k+1},u_i\rangle$ or $\langle u_i,w\rangle$ belongs to $\delta^+(S)$ for each $1\leq i\leq k$; thus, $x^*(\delta^+(S))\geq 1$. If $w\in S$, either $\langle w,v_i\rangle$ or $\langle v_i,v_0\rangle$ belongs to $\delta^+(S)$ for every $i$ and therefore $x^*(\delta^+(S))\geq 1$.
\end{proof}

Let $x'$ be an arbitrary solution to the linear system we will construct; our goal is establishing $x'=x^*$.

For every $(p,q)\in V\times V$ such that $x^*_{pq}=0$, we choose the tight constraint $x_{pq}\geq 0$: i.e., $x_{pq}=0$ is added to the linear system. This implies that $x'$ will be supported on (a subset of) the support of $x^*$, and therefore we can now focus on the edges in the support of $x^*$ only.

Let $S_i:=V\setminus\{u_i,.\ldots,u_{k+1}\}$ for $1\leq i\leq k+1$. For each $i$, we choose the tight constraint $x(\delta^+(S_i))\geq 1$; this implies that $x'(\delta^+(S_i))=x'_{u_0u_{k+1}}+x'_{u_{i-1}u_i}=1$, from the earlier observation that $x'$ is supported by a subset of the support of $x^*$. Thus, we have \begin{equation}\label{e:b1}
x'_{u_0u_1}=x'_{u_1u_2}=\cdots=x'_{u_ku_{k+1}}
.\end{equation}By choosing the outdegree constraints of $u_1,\ldots,u_k$, we have $1=x'_{u_1u_2}+x'_{u_1 w}=x'_{u_2u_3}+x'_{u_2 w}=\cdots =x'_{u_ku_{k+1}}+x'_{u_k w}$ and therefore $x'_{u_1 w}=\cdots=x'_{u_k w}$ from \eqref{e:b1}. Then choosing the indegree constraint of $w$ gives $x'_{u_1 w}=\cdots=x'_{u_k w}=\frac{1}{k}$. Similarly, by choosing the indegree constraints of $u_1,\ldots,u_k$, we have $x'_{u_{k+1}u_1}=x'_{u_{k+1}u_2}=\cdots=x'_{u_{k+1}u_k}$, and the outdegree constraint of $u_{k+1}$ gives $x'_{u_{k+1}u_1}=x'_{u_{k+1}u_2}=\cdots=x'_{u_{k+1}u_k}=\frac{1}{k}$. Now the indegree constraint of $u_1$ gives $x'_{u_0u_1}=1-\frac{1}{k}$ which in turn shows $x'_{u_1u_2}=\cdots=x'_{u_ku_{k+1}}=1-\frac{1}{k}$ and the outdegree constraint of $u_0$ yields $x'_{u_0u_{k+1}}=\frac{1}{k}$. By symmetry, we have $x'_{pq}=x^*_{pq}$ for $p,q\in\{w,v_0,\ldots,v_{k+1}\}$ as well.

Finally, by choosing the tight constraint $x(\delta^+(V\setminus\{u_0\}))\geq 1$, we have $x'_{v_{k+1}u_0}=1$. This completes the proof that $x'=x^*$.
\end{proof}

\section*{Acknowledgments.} The authors thank the anonymous reviewers of a conference version of this paper~\cite{AKS:conf} for their helpful comments.

\end{document}